\documentclass[twocolumn,showpacs,preprintnumbers,amsmath,amssymb]{revtex4}

\usepackage{graphicx}
\usepackage{dcolumn}
\usepackage{bm}
\usepackage{subfigure}

\def\lapproxeq{\lower .7ex\hbox{$\;\stackrel{\textstyle
<}{\sim}\;$}}
\def\gapproxeq{\lower .7ex\hbox{$\;\stackrel{\textstyle
>}{\sim}\;$}}

\begin{document}



\title{
New facts about muon production in Extended Air Shower simulations
}

\author{
T.~Pierog$^{1}$}
\email{tanguy.pierog@ik.fzk.de}
\author{
K.~Werner$^2$}
\email{werner@subatech.in2p3.fr}
\affiliation{
(1) Forschungszentrum Karlsruhe, Institut f\"ur Kernphysik, Karlsruhe, Germany\\
(2) SUBATECH, University of Nantes -- IN2P3/CNRS-- EMN,  Nantes, France
}

\date{\today}

\begin{abstract}

Whereas air shower simulations are very valuable tools for interpreting cosmic ray data,
there is a long standing problem: is seems to be impossible to accommodate at the
same time the longitudinal development of air showers and the 
number of muons measured at ground. Using a new hadronic interaction model (EPOS) in
air shower simulations produces considerably more muons, in agreement with results
from the HiRes-MIA experiment. We find that this is mainly due to a better description of
baryon-antibaryon production in hadronic interactions. This is a new aspect of air shower
 physics which has never been considered so far.

\end{abstract}



\maketitle

Since more than ten years detailed extended air shower (EAS) simulations play a 
decisive role in
interpreting measurements from ground based cosmic ray measurements. This concerns for
example the chemical composition of cosmic rays in the KASCADE experiment~\cite{kascade} 
or the primary energy determination in AGASA array~\cite{agasa}.

An air shower is initiated by a very  energetic proton or nucleus (primary particle),
which interacts with air by producing many secondary hadrons, which interact again, and so on.
Neutral pions play a special role, since they decay into gammas which initiate an
electromagnetic shower each. The latter one is well under control (elementary processes of
QED), whereas the hadronic interactions require models, being tested against accelerator data -- 
at much lower energies though than the highest primary energies. It turns out that more
than 90\% of the initial energy goes into the (well-known) electromagnetic part, 
whereas the rest shows up as muons from the hadronic decays. The muonic part depends
therefore strongly on the hadronic modeling (up to a factor of 2 difference in different models), 
while the longitudinal development of the electromagnetic part (and in particular
 its maximum $\mathrm{X}_{\mathrm{max}}$) are relatively robust (less than 10\% variations
between models).

Using the currently employed hadronic interaction models (QGSJET01~\cite{qgsjet01}, 
QGSJET~II~\cite{qgsjet2}, SIBYLL~2.1~\cite{sibyll2.1}), one has a tendency to have less
muons in the simulations than observed by the experiments. Direct measurements of 
high energy muons (\~100 TeV)~\cite{cosmoaleph, delphi, l3+c,amanda} as well as experiments 
using low energy muons (\~1 GeV) like KASCADE~\cite{kascade} or HiRes-MIA~\cite{mia}, show inconsistencies 
between experimental data
and simulations. Furthermore, at very high energy, the Pierre Auger Observatory finds
 a discrepancy between the energy reconstruction of the primary
cosmic rays using a purely experimental method or using a method based partly on the muon 
density at ground~\cite{Sommers:2005vs} and air shower simulations. The 25\%
difference could be explained by a lack of muons in the simulations.
Many attempts have been made to force the models to increase the muon production
without changing $\mathrm{X}_{\mathrm{max}}$ (well constrained by data)
without success~\cite{hoerandel}. 

It should be noted that the number of muons predicted from shower simulations
 has important consequences on the astrophysical interpretation 
of the very high energy cosmic ray spectrum. If the muon number currently 
predicted is right,  ``new physic'' has to be 
introduced to explain the observation of events beyond the ``GZK-cutoff''~\cite{gzk} of 
$10^{20}$eV, observed  by the AGASA experiment (for a review see~\cite{bhat-sigl}).

In this work, we discuss the consequences of introducing the recently developed 
high energy hadronic interaction 
model EPOS into the air shower simulation models CORSIKA~\cite{corsika} 
and CONEX~\cite{conex0,conex,Bergmann:2006yz}. EPOS is a consistent quantum mechanical 
multiple scattering approach 
based on partons and strings~\cite{Drescher:2000ha}, where cross sections and 
the particle production are calculated 
consistently, taking into account energy conservation in both cases (unlike
other models where energy conservation is not considered for cross section 
calculations~\cite{Hladik:2001zy}). 
A special feature is the explicit treatment of projectile and target remnants,
leading to a very good description of baryon and antibaryon production as measured
in proton-proton collisions at 158~GeV at CERN~\cite{Liu:2003wj}.
Motivated by the very nice data obtained by the RHIC experiments, nuclear effects
 related to CRONIN transverse momentum broadening~\cite{Cronin:zm}, parton saturation, 
 and screening have been introduced into
EPOS~\cite{Werner:2005jf}. Furthermore high density effects  leading to collective 
behavior in heavy ion collisions are also taken into account~\cite{Werner:2006wp}. 
It appears that EPOS does very well
compared to RHIC data~\cite{Bellwied:2005bi,Abelev:2006cs}, and also
all other available data from high energy particle physic experiments 
(ISR,CDF and especially SPS experiments at CERN)~\cite{Werner:2006xxx}. As a result, it is the only
model used both for EAS simulations and accelerator physics which is able to 
reproduce consistently almost all data from 100 GeV lab to 1.8 TeV center of mass energy, 
including antibaryons, multi-strange particles, ratios and pt distributions. 
It is intensively tested against data, but as explained, it is not a simple fit of data.
And in particular, since this model is used for accelerator physics, many data are used 
to constrain the model parameters which are not a priori linked to cosmic rays and 
air showers.

For our analysis, CONEX and CORSIKA are used to simulate the air shower development, 
using GHEISHA~\cite{gheisha} as low energy hadronic interaction model below
80 GeV. For the high energy interactions (above 80 GeV) EPOS 1.35 is used, and as a 
reference the most commonly used interaction model QGSJET01.  

\begin{figure}
\centerline{
\includegraphics[width=8.2cm]{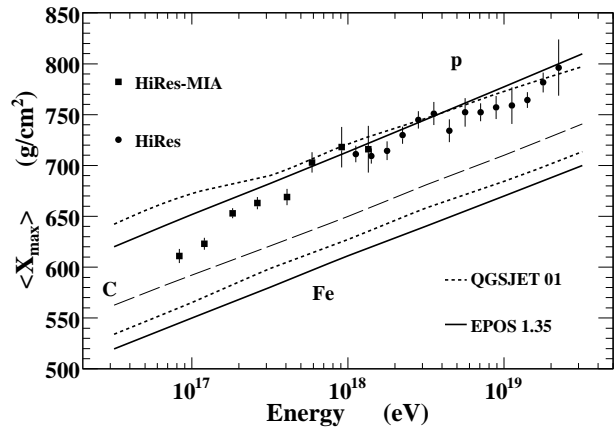}
}
\caption{The shower maximum $\mathrm{X}_{\mathrm{max}}$ as measured by the HiRes-MIA~\cite{mia} and the HiRes
collaboration~\cite{hires} as a function of primary energy,
compared to proton and iron induced showers simulated
with EPOS (full lines) and QGSJET01 (dotted lines) as high energy hadronic interaction model.
The thin line is calculated
with  EPOS  for carbon nuclei as primaries.}
\label{fig-xmax}
\end{figure}

One of the most important observables in air shower physics is 
the distribution of the electron number as a function of the depth $\mathrm{X}$, 
the latter one representing the amount of air traversed  by the shower, 
expressed in g/cm$^2$. The maximum $\mathrm{X}_{\mathrm{max}}$ of this distribution 
is a function of the energy of the primary particle, as shown in Fig. \ref{fig-xmax}, 
where the results represent averages over many showers. The experimental data
from HiRes~\cite{hires} (points in Fig. \ref{fig-xmax}) refer to unknown primary particles, therefore 
one usually compares the data with the two extremes (protons and iron) from
simulations. Here we show results for EPOS (full lines) and QGSJET01 
(dotted lines). The upper lines represent protons and the lower lines iron.
 Both models are compatible
with experimental data which seems to show a lightening of the primary cosmic ray
composition between $10^{17}$~eV and $10^{18}$~eV. But even at the lower energies, 
the average primary particle
doesn't seem to be heavier than a carbon nucleus as shown by the dashed line
calculated with EPOS for carbon-induced showers. Above  $10^{18}$~eV, data are well 
reproduced by proton induced shower. 

\begin{figure}
\centerline{
\includegraphics[width=8.2cm]{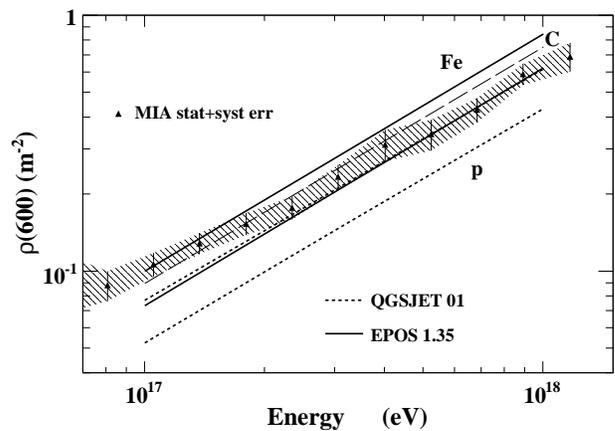}
}
\caption{
The muon density $\rho_\mu(600)$ from the MIA experiment~\cite{mia} (triangles) as a function of
 the primary energy, compared to simulated
proton and iron induced showers with EPOS (full line) and QGSJET01 (dotted line)
as high energy hadronic interaction model. For better comparison with 
Fig.~1, the  dashed line correspond to carbon induced showers with EPOS.}
\label{fig-mudens}
\end{figure}

A complementary observable is the muon number at ground, for example expressed
via the density $\rho_\mu(600)$ of muons per squared meter 
at a lateral distance of 600 m from the shower core (impact point) as measured by the 
MIA~\cite{mia} detector as a function of the primary energy, shown in  Fig. \ref{fig-mudens} 
as triangles.
Again we show shower simulations for protons and iron for EPOS and QGSJET01, 
using the same conventions as in Fig.~\ref{fig-xmax}. 
The  HiRes-MIA data are  now compatible with the EPOS results,
using a heavier primary composition (but not too heavy, more like carbon) at 
$10^{17}$~eV and a lighter one
(proton) at $10^{18}$~eV. Compared to QGSJET01, it is a shift of about
 40\% in the number of muons at ground. The proton line from EPOS lies exactly on
the iron line from QGSJET01 (which was the model giving the highest number of muons
before). 

So for the first time, both $\mathrm{X}_{\mathrm{max}}$ and muon data are compatible 
with a change of the average incident particle from carbon to proton, between 
  $10^{17}$~eV and $10^{18}$~eV.

\begin{figure}
\centerline{
\includegraphics[width=8.0cm]{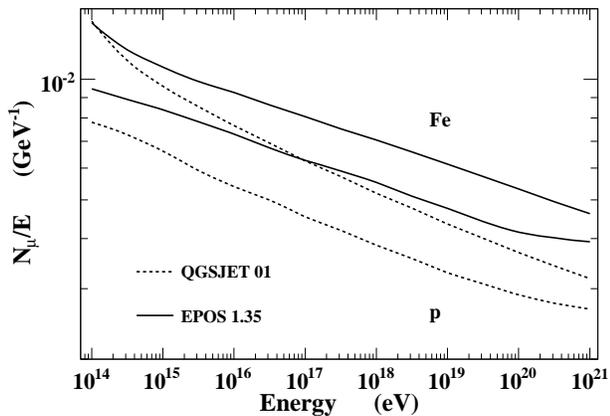}
}
\caption{Total number of muons at ground divided by the primary energy expressed in GeV
as a function of the primary energy for proton and iron induced shower using QGSJET01
(dotted lines) or EPOS (full lines) as high energy interaction model.}
\label{fig-ratio}
\end{figure}

Not only the absolute value of the muon density has changed but also the slope
is slightly higher (less negative) in EPOS compared to QGSJET01. This can be seen on a larger energy
scale Fig. \ref{fig-ratio}: the number of muons arriving at ground divided by the primary 
energy in GeV
is shown as a function of the primary energy between $10^{14}$ eV and $10^{21}$ eV.
The EPOS curves are much flatter:
at the lowest energy, the EPOS proton line
is at most 25\% higher than QGSJET, but at $10^{20}$ eV EPOS is a factor 2 higher and
gives even more muons with a  primary proton than QGSJET01 for iron induced showers. As
a result, EPOS should be compatible with the KASCADE data. It will most probably give a
 lighter average composition, but a precise answer has to await full simulations (in preparation).
However, EPOS will give very different results for the AUGER spectrum based on
surface detector for 
instance compared to QGSJET01 simulations.
In addition, EPOS has a slightly higher elongation rate (slope of 
$\mathrm{X}_{\mathrm{max}}$ as a function of primary energy) compared to QGSJET01 but 
stays fully compatible with the HiRes data at highest energies.


For a deeper understanding of what makes the difference between EPOS and older 
models, one has to have a closer look at the EAS physics. Looking
at all particles produced in EAS by EPOS, we find that not only
the number of muons is increased but also the number of baryons and antibaryons 
 -- referred to as (anti)baryons in the following.
As a test, we modified EPOS parameters artificially to reduce the number of produced (anti)baryons 
by  a factor of two, hence not reproducing 
accelerator data anymore. 
As a result, the number of muons
is reduced on the average by 30\%, being quite close to QGSJET01. At the same time,
$\mathrm{X}_{\mathrm{max}}$ is not changed.
So we find -- and this has never been discussed in the literature so far -- 
that (anti)baryon production is a very efficient mechanism to affect 
the muon numbers without touching  $\mathrm{X}_{\mathrm{max}}$.

As a cross
check, we increased artificially the (anti)baryon production in  the SIBYLL~\cite{sibyll}
hadronic interaction mode (commonly used for 
EAS simulations) by a factor of 2, resulting in 30\%  more muons.

Up to now, the correlation between the number of (anti)baryons (about 1\% of the particles in the
hadronic shower) and the number of muons in air showers has never been studied. 
But, thanks to a simple 
Heitler model generalized to hadronic showers \cite{heitler,matt}, one can easily 
understand the role of the 
antibaryons in EAS. In this kind of toy model, a hadronic interaction of a charged 
particle  with energy $E$ will produce $N_{\rm tot}$ new particles with 
energy $E/N_{\rm tot}$, with $N_{\rm EM}$  particles (mainly $\pi^0$) 
transferring their energy to the electromagnetic channel 
 via immediate decay into  photons, each one initiating an  EM shower. 
The other $N_{\rm had}=N_{\rm tot}-N_{\rm EM}$ particles re-interact with air
 after a constant interaction path length $\lambda_{\rm had}$ 
 and thus contribute to the hadronic cascade. 
In the electromagnetic shower, exactly 2 secondaries with equally
shared energy are produced at each interaction,
after a constant interaction path length $\lambda_{\rm EM}$. 
Introducing a characteristic energy 
($E_{\rm dec}=150$~GeV), where pions are assumed to decay into muons,
the number of muons for a shower with primary energy $E_0$ can be written as \cite{Pierog:2006qu}
\begin{equation}
N_{\mu}=\{N_{\rm had}\}^n=\left( \frac{E_0}{E_{\rm dec}} \right) ^ \alpha,
\end{equation} 
with $\alpha=\ln{N_{\rm had}}/\ln{N_{\rm tot}}<1$, and
where $n$ is the number of hadronic generations in the shower. 
Introducing 
$R=N_{\rm had}/N_{\rm tot}$, we have
\begin{equation}
\alpha=1+\frac{\ln{R}}{\ln{N_{\rm tot}}}.
\label{eq-alpha}
\end{equation} 
Eq. (\ref{eq-alpha}) shows that the muon number 
depends strongly on the ratio $R$ of the number hadrons initiating a hadronic sub-cascade to the total 
multiplicity, which is understandable since the difference between these two quantities are particles
giving all their energy to the electromagnetic channel -- not producing muons.

Usually these kind
of toy models consider only pions as secondary particles. As a consequence, one has $R=2/3$. 
So the muon number depends only on  $N_{\rm tot}$. The latter one affects as well  $\mathrm{X}_{\mathrm{max}}$, 
since from our toy model we obtain~\cite{Pierog:2006qu}
\begin{equation}
X_{\rm max} = \lambda_{\rm had}+\lambda_{\rm EM}\cdot\ln\left( \frac{E_0}{N_{\rm tot}E_c}\right),  \label{eqxmax}
\end{equation} 
where  $E_c=85$ MeV is the critical
energy (energy of the electromagnetic particles at the shower maximum in air and energy
of particles disappearing from the shower). 
So EAS simulations based on two different interaction models, producing different
total multiplicities, should disagree for both $X_{\rm max}$  and muon numbers.

Let us now be  more realistic, and consider all kinds of hadrons, including (anti)baryons.
Particle production in hadronic interactions is model dependent, and so is the precise 
value of $R$.
With $R$ being less than 1 and $N_{\rm tot}>>1$,   the muon number depends very sensitively 
on the ratio $R$.
One may imagine two interaction models with the same $N_{\rm tot}$ but with more  baryons in 
one model compared to the other, corresponding to  smaller $N_{\rm EM}$ and bigger $R$ values.
With $R$ being even slightly 
higher, $\alpha$ is closer to one, increasing  both the number of muons and the slope as a 
function of the energy as observed Fig. \ref{fig-ratio}.
This muon increase is also intuitively understandable: 
more baryons in one model compared to the other lead to  a smaller $N_{\rm EM}$,  
less energy is transfered to the electromagnetic component. There are more
hadronic generations and thus more muons are produced. 

\begin{figure}
\centerline{
\includegraphics[width=8.2cm]{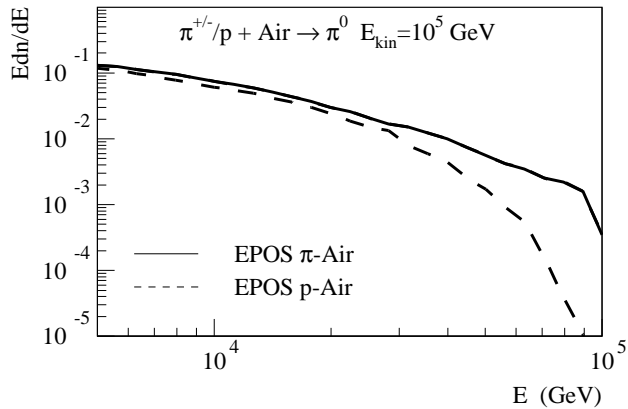}
}
\caption{EPOS $\pi^0$ energy spectra for $\pi$-$Air$ interactions (full line) 
and for $p$-$Air$ interactions (dashed line) at $10^5$~GeV kinetic energy.}
\label{fig-spectra}
\end{figure}

As a result, it is clear that variations of the number of baryons and antibaryons
affect $R$, but why is the effect so strong although baryons and antibaryons represent
only 1\% of all hadrons. 
One should not forget that only hadrons with a large longitudinal 
momentum fraction $x_E$ ($>$0.1) contribute significantly to the cascade.
 Particle yields at large $x_E$ 
are quite different from those at central rapidities, measured in accelerator experiments.  
Fig. \ref{fig-spectra} shows the energy spectra of $\pi^0$ produced in $p$-$Air$ (dashed line) and $\pi$-$Air$ (full line)
reactions at $10^5$~GeV kinetic energy for EPOS. 
 $\pi^0$'s produced in $p$-$Air$ interactions are 
much softer than in $\pi$-$Air$ simply because
$p$-$Air$ interactions do not produce leading $\pi^0$ whereas $\pi$-$Air$ do so because of
charge exchange. As a consequence, the ratio  $R \approx 1-N_{\pi^0}/N_{\rm tot}$ for particles 
with more than 50\% of the interaction energy is close to 15\% for $\pi$-$Air$ but
about 1\% for $p$-$Air$. In other words, producing only slightly more baryons per hadronic interaction
in one model compared to another 
will decrease considerably the number of fast neutral pions ($\approx N_{\rm EM}$) in 
the next generation, leaving more contributors to the hadronic cascade and providing thus more muons. 
It is worthwhile to say that having by any means more leading
 $\pi^0$, will decrease the number of muons and the slope as a 
function of the energy~\cite{Alvarez-Muniz:2006ye}.

Why does the muon yield in EPOS differ so much from other models? 
As already said, EPOS has not been designed to be used for EAS simulations, but to
understand hadronic interactions. Therefore  all available data on
p-p, p-A, A-A, $\pi$-p/A, K-p/A have been considered, in particular 
data on observables like (anti)baryons which so far have not been considered 
to be important for EAS simulations.
Differences between hadronic
interaction models used for EAS simulation will be discussed in detail in a 
forthcoming publication.

To summarize: simulating air showers by using the new high energy hadronic interaction model EPOS
results in an increase of the muon density at $10^{18}$~eV of about 40\% compared to QGSJET01 calculations.
So for the first time, both $\mathrm{X}_{\mathrm{max}}$ and muon data are compatible 
with a change of the average incident particle from heavy to light element, in the energy range between 
  $10^{17}$~eV and $10^{18}$~eV.
It was shown for the first time that (anti)baryon production
plays a much more important role in EAS physics than expected. 
An increased (anti)baryon production (in one model compared to another)
increases  the number of interactions where no leading  $\pi^0$ is produced, 
more energy goes into hadronic sub-showers, leading to more hadron generations,
and finally to more muons.

The presented EPOS results have certainly to be confirmed by comparing to other air
shower experiments (work in progress), but it seems already now that the Cosmic Ray
energy spectrum based on EAS simulation will have to rescale its energy to lower
values. Another consequence:  since there is more energy in the hadronic
part of the shower, the calorimetric energy measured by fluorescence detectors such as
HiRes will correspond to a larger primary energy. The conversion factor is changed by 
about 5\%. A more quantitative statement on the  CR spectrum requires intensive simulations
which will be presented soon.
 
{\it Acknowledgments.}
The authors are grateful to R.~Engel, D.~Heck and S.S.~Ostapchenko for very helpful 
discussions.


\end{document}